\documentclass{aa}

\usepackage[varg]{txfonts}

\usepackage{color,amsmath}

\bibpunct{(}{)}{;}{a}{}{,} % to follow the A&A style

\newcommand{\edit}[1]{#1}
\newcommand{\Ab}{{\boldsymbol A}}
\newcommand{\Bb}{{\boldsymbol B}}
\newcommand{\Eb}{{\boldsymbol E}}
\newcommand{\eb}{{\boldsymbol e}}
\newcommand{\jb}{{\boldsymbol j}}
\newcommand{\lb}{{\boldsymbol l}}
\newcommand{\nb}{{\boldsymbol n}}
\newcommand{\vb}{{\boldsymbol v}}
{

\title{The global distribution of magnetic helicity in the solar corona}

\author{A. R. Yeates\inst{\ref{inst1}}
\and G. Hornig\inst{\ref{inst2}}
}

\institute{Department of Mathematical Sciences, Durham University, Durham, DH1 3LE, UK \email{anthony.yeates@durham.ac.uk}\label{inst1}
\and
Division of Mathematics, University of Dundee, Dundee, DD1 4HN, UK \label{inst2}
}

\date{Received ?? / Accepted ??}

\abstract{
By defining an appropriate field line helicity, we apply the powerful concept of magnetic helicity to the problem of global magnetic field evolution in the Sun's corona. As an ideal-magnetohydrodynamic invariant, the field line helicity is a meaningful measure of how magnetic helicity is distributed within the coronal volume. It may be interpreted, for each magnetic field line, as a magnetic flux linking with that field line.  Using magneto-frictional simulations, we investigate how field line helicity evolves in the non-potential corona as a result of shearing by large-scale motions on the solar surface. On open magnetic field lines, the helicity injected by the Sun is largely output to the solar wind, provided that the coronal relaxation is sufficiently fast. But on closed magnetic field lines, helicity is able to build up. We find that the field line helicity is non-uniformly distributed, and is highly concentrated in twisted magnetic flux ropes. Eruption of these flux ropes is shown to lead to sudden bursts of helicity output, in contrast to the steady flux along the open magnetic field lines.
}


\begin{document}

\maketitle

\section{Introduction}

Magnetic helicity is well-known to be invariant in ideal magnetohydrodynamics (MHD), and in highly conducting plasmas it is almost conserved even for finite resistivity  \citep{berger1984,pariat2015}. Helicity may be interpreted as a net linking or winding of magnetic field lines around one another \citep{moffatt1969}. This linking can put a lower bound on the magnetic energy \citep{moffatt1990,Freedman1991,Berger1993b}, reflecting the physical barrier that, in ideal MHD, magnetic field lines are unable to pass through one another or to reconnect.

In the corona, which is highly conducting and close to ideal, there are two primary sources of helicity, both coming from the solar interior: shearing of magnetic field lines by footpoint motions, and emergence of twisted magnetic fields. In this paper, we will consider only the production of helicity by footpoint motions. These footpoint motions may arise either from large-scale flows (primarily differential rotation), or from small-scale convection. Here we model the net effect of convection on the large-scale magnetic field with an isotropic diffusion on the solar surface. We neglect the additional helicity injection that could arise if the convective motions had a net sign of vorticity \citep{antiochos2013,mackay2014,knizhnik2015}.

Perhaps the most important practical consequence of helicity in the corona is the formation of twisted magnetic flux ropes, and their eruption as coronal mass ejections \citep[e.g.,][]{chen2011}. These remove helicity from the corona and send it out into the heliosphere. However, a difficulty in quantifying this notion arises because helicity is a volume integral, and is not conserved on an arbitrary sub-volume of the corona. Previous authors have quantified the helicity generated by solar rotation in two extreme cases: entire hemispheres \citep{berger2000}, and a single isolated active region \citep{devore2000}. The goal of this paper is to show how one can meaningfully study the spatial distribution of helicity within the corona, with the ultimate aim of understanding the origin of solar eruptions. The basic idea is to decompose the corona into infinitesimal tubular volumes around each magnetic field line. The helicity of each of these sub-volumes is an ideal invariant (provided that the field line endpoints are fixed), called the field line helicity. 

This idea of field line helicity is not a new one, and goes back to \citet{Taylor1974s}. Subsequently, \citet{berger1988} derived lower energy bounds based on field line helicity \citep[see also][]{aly2014,yeates2014c}, but the concept was not significantly developed for a number of years. More recently, field line helicity has been found to be an invaluable tool for understanding the turbulent relaxation of braided magnetic fields. For cylindrical domains, \citet{yeates2013,yeates2014b} proved that knowing the field line helicity on each field line uniquely determines the field line mapping from one end of the cylinder to the other. And in sufficiently complex magnetic fields, \citet{russell2015} showed that field line helicity is efficiently redistributed by reconnection, but is not destroyed on dynamical timescales. It therefore acts as a constraint on magnetic relaxation, demonstrated in the numerical experiments of \citet{pontin2011}.

The aim of this paper is to apply the tool of field line helicity to the Sun's corona, in which the magnetic field has a rather more complex topology than the cylinder. Accordingly, the "completeness" proof of \citet{yeates2014b} no longer applies, though field line helicity retains its importance as a topological invariant.

Since the global magnetic field in the coronal volume can not be measured directly, a numerical model is required. However, for the study of helicity, a model more sophisticated than potential field extrapolation is needed. Primarily, this is because potential extrapolations lack the free magnetic energy that is associated with helicity. But it is also because they do not evolve continuously over time, so do not preserve the connectivity of magnetic field lines associated with an ideal evolution. In other words, a sequence of potential field extrapolations could ``undo'' the field line entanglement imposed in reality by the footpoint motions. Instead, in order to model the gradual injection of helicity over time, a time dependent model is required. Here we apply the magneto-frictional model \citep{vanballegooijen2000}, as a compromise that retains sufficient physics but is less computationally expensive than full-MHD simulations. The importance of retaining a continuous time dependence has been demonstrated before, but will be shown rather clearly by the field line helicity.

The paper is organised as follows. Section \ref{sec:flh} explains the physical interpretation of field line helicity, and gives a practical definition, then Sect. \ref{sec:mf} describes the magneto-frictional model. We then study three situations of increasing complexity: a dipolar field (Sect. \ref{sec:dipolar}), a quadrupolar field (Sect. \ref{sec:quadrupolar}), and finally a more realistic, non-axisymmetric configuration (Sect. \ref{sec:nonaxi}). We conclude in Sect. \ref{sec:conc}.

\section{Field line helicity} \label{sec:flh}

We model the solar corona by a spherical shell $D = \{(r,\theta,\phi)\,|\, r_0 < r < r_1 \}$. The field line helicity of a magnetic field line $L\subset D$ is defined as
\begin{equation}
{\cal A}(L) := \int_{L(x)}{\Ab}\cdot\eb_B\,\mathrm{d}l,
\end{equation}
where $\eb_B = {\Bb}/|{\Bb}|$ is the unit vector aligned with the local direction of the magnetic field ${\Bb}=\nabla\times{\Ab}$. It follows that ${\cal A}$ is undefined on ergodic magnetic field lines, which have infinite length. In generic coronal magnetic fields, this situation does not usually arise since field lines are typically finite in length, ending on one or more of the domain boundaries $r=r_0$ and $r=r_1$. The choice of $\Ab$ will be discussed below.

Since there is a unique field line through each point (except for magnetic null points where $\Bb=0$), we can also assign values of ${\cal A}$ to  points $x\in D$, and think of ${\cal A}$ as a function on $D$. This can be useful for visualization. This function ${\cal A}$ is evidently constant along magnetic field lines, and will, furthermore, be continuous in regions of continuous field line mapping. In the presence of magnetic null points, ${\cal A}$ will generally be discontinuous across their separatrix surfaces, like any field-line integrated quantity.

Although we have defined ${\cal A}$ as a line integral, it may also be written as the limit
\begin{equation}
{\cal A}(L) = \lim_{\epsilon\to 0}\frac{1}{\Phi_\epsilon}\int_{D_\epsilon}\Ab\cdot\Bb\,\mathrm{d}V,
\label{eqn:lim}
\end{equation}
where $D_\epsilon$ is the magnetic flux tube of radius $\epsilon$ around the field line $L$, with $\Phi_\epsilon$ being the flux of this tube. This motivates the name ``field line helicity'' \citep{berger1988}. Integrating \eqref{eqn:lim} over all field lines, weighted by their flux, will recover the total helicity $H=\int_D\Ab\cdot\Bb\,\mathrm{d}V$. In this sense, ${\cal A}$ is a meaningful density for $H$, describing how topological sub-structure is distributed within $D$.

\subsection{Physical interpretation as magnetic flux} \label{sec:flux}

\begin{figure}
\resizebox{\hsize}{!}{\includegraphics{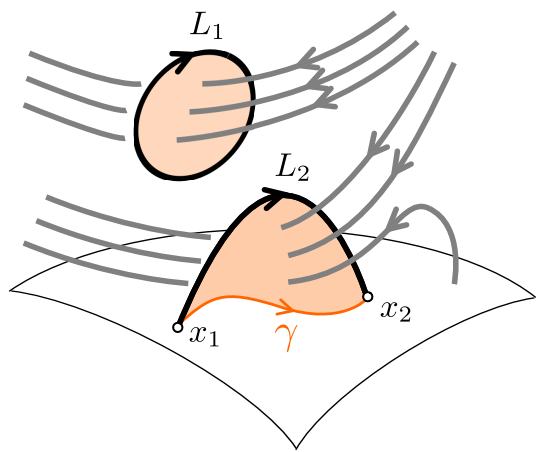}}
\caption{Physical interpretation of field line helicity ${\cal A}$ as the magnetic flux linking a closed ($L_1$) or open ($L_2$) magnetic field line.}
\label{fig:flux}
\end{figure}

The physical meaning of ${\cal A}$ is clear when $L$ is a closed curve such as $L_1$ in Fig. \ref{fig:flux}. In that case, Stokes' theorem implies that ${\cal A}$ is simply the magnetic flux that links through $L$. It follows that ${\cal A}$ must be an ideal invariant when $L$ is a closed curve. This conclusion does not depend on the chosen gauge of $\Ab$, and indeed the value of ${\cal A}$ is independent of this gauge (provided that $\Ab$ is single-valued).

In the coronal situation, closed magnetic field lines are rare, and we must consider field lines ending on one or more of the domain boundaries ($r=r_0$ and $r=r_1$). For example, consider the field line $L_2$ in Fig. \ref{fig:flux}, linking two points $x_1$ and $x_2$ on $r=r_0$. Under a gauge transformation from $\Ab$ to $\Ab' = \Ab + \nabla\chi$, the field line helicity changes from ${\cal A}(L_2)$ to ${\cal A}'(L_2) = {\cal A}(L_2) + \chi(x_2) - \chi(x_1)$. However, whichever gauge is used, we can always find a corresponding surface $S$ whose magnetic flux is exactly ${\cal A}(L_2)$. This is equivalent to the existence of some curve $\gamma$ between $x_1$ and $x_2$ (as in Fig. \ref{fig:flux}) such that $\int_\gamma\Ab\cdot\,\mathrm{d}\lb=0$, which is demonstrated in Appendix A.

In different gauges, the possible curves $\gamma$ will differ, so that the physical fluxes represented by ${\cal A}$ will depend on the chosen gauge. But, whichever gauge is chosen, we may interpret ${\cal A}$ in terms of the physical linking of fluxes. Note that it is only the gauge on the boundary $\partial D$ that matters; this is analogous to the situation with relative helicity \citep{berger1984a, prior2014}. For field lines such as $L_2$, \citet{antiochos1987} defined a ``flux-per-field-line'' which is equivalent to ${\cal A}$ if we take $\gamma$ to be a geodesic on $\partial D$ between $x_1$ and $x_2$. However, rather than specifying the curve $\gamma$ explicitly like this, we will specify the gauge of $\Ab$ explicitly, as described in the next section. This makes ${\cal A}$ is easier to compute, although it will lead (in general) to more complex $\gamma$.

The interpretation in terms of linked fluxes shows that ${\cal A}$ remains an ideal invariant in the case of non-closed field lines, provided that the gauge is fixed and, in addition, that there are no motions of field line footpoints on $\partial D$. Of course, an important feature of the corona is the injection of helicity through footpoint motions, which will necessarily lead to a change in ${\cal A}$. This will be demonstrated in Sects. \ref{sec:dipolar}--\ref{sec:nonaxi}.

\subsection{Gauge choice} \label{sec:gauge}

For practical application, we must choose a gauge for $\Ab$. In this paper, we use the so-called DeVore gauge, chosen since it is straightforward to compute and leads to a clear physical interpretation for ${\cal A}$. The gauge was introduced by \citet{devore2000} in \edit{Cartesian} coordinates, and has been used by a number of authors in both \edit{Cartesian} geometry \citep{valori2012, moraitis2014} and spherical geometry \citep{amari2013}.

The DeVore gauge condition is that $A_r\equiv 0$. From $\Bb=\nabla\times\Ab$, we get
\begin{equation}
\Bb\times\eb_r = \frac{1}{r}\frac{\partial}{\partial r}\left(r\Ab\right),
\end{equation}
which may be integrated in $r$ to give
\begin{equation}
r\Ab(r,\theta,\phi) = r_0\Ab_0(\theta,\phi) + \int_{r_0}^{r}\Bb(r',\theta,\phi)\times\eb_r\,r'\,\mathrm{d}r'.
\label{eqn:devore}
\end{equation}
Here $\Ab_0$ is the vector potential on the initial surface $r=r_0$. Under our assumption that $A_r\equiv 0$, it follows that $\Ab_0$ has only $\theta$ and $\phi$ components, which must satisfy
\begin{equation}
\eb_r\cdot\nabla\times\Ab_0 = B_r(r_0,\theta,\phi),
\label{eqn:br}
\end{equation}
but are otherwise arbitrary. We will follow \citet{amari2013} and fix $\Ab_0$ with the  condition $\nabla\cdot\Ab_0=0$, \edit{so that it may be written as}
\begin{equation}
\Ab_0(\theta,\phi) = \nabla_\perp\psi(\theta,\phi)\times\eb_r.
\label{eqn:A0}
\end{equation}
\edit{Note that $\psi$ is (up to a constant) the poloidal flux function from a poloidal-toroidal decomposition of $\Bb$. Equation \eqref{eqn:br} then} requires that
\begin{equation}
\nabla_\perp^2\psi = -B_r(r_0,\theta,\phi).
\label{eqn:poisson}
\end{equation}
Solving this Poisson equation for $\psi$ on the sphere determines $\Ab_0$. The function $\psi$ is determined only up to an additive constant, which we may fix by requiring that $\int_{r=r_0}\psi\,\mathrm{d}\Omega=0$. In spherical geometry, the Green's function for \eqref{eqn:poisson} is known, and the solution may be expressed analytically \citep{kimura1987} as
\begin{equation}
\psi(\theta,\phi) = \frac{1}{4\pi}\int_{r=r_0}B_r(\theta',\phi')\log\big(1 - \cos\xi\big)\sin\theta'\,\mathrm{d}\theta'\,\mathrm{d}\phi',
\label{eqn:psi}
\end{equation}
where
\begin{equation}
\cos\xi = \cos\theta\cos\theta' + \sin\theta\sin\theta'\cos(\phi-\phi').
\end{equation}

In the special case of a potential field $\Bb=\Bb_{\rm p}$ (i.e., $\nabla\times\Bb_{\rm p}=0$), it follows from \eqref{eqn:devore} that $\nabla\cdot\Ab_{\rm p}=0$ everywhere in $V$. For a potential field in this (Coulomb) gauge, we have $\int_D\Ab_{\rm p}\cdot\Bb_{\rm p}\,\mathrm{d}V=0$ \citep{berger1984}, although ${\cal A}$ need not vanish for any individual field line -- we will see an example of this in Sect. \ref{sec:nonaxi}.

\subsection{Physical interpretation of the gauge choice}

One advantage of the DeVore gauge \eqref{eqn:devore} is its explicit physical interpretation. We consider the contributions to ${\cal A}$ from both the integral term in \eqref{eqn:devore} and the boundary term $r\Ab_0$.

\begin{figure}
\resizebox{\hsize}{!}{\includegraphics{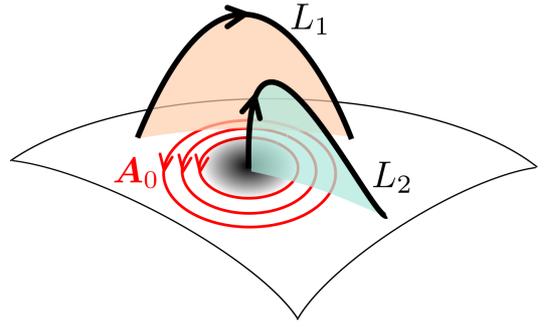}}
\caption{Physical interpretation of field line helicity in the DeVore gauge. The red circles show the direction of $\Ab_0$ (contours of $\psi$) arising from a strong magnetic source $B_r>0$. The shaded surfaces are radial projections of the field lines $L_1$ and $L_2$. Both field lines have a contribution to ${\cal A}$ from any flux linking through these surfaces (owing to the second term of \eqref{eqn:devore}), but only $L_1$ has a contribution from $\Ab_0$, since the projection of $L_2$ is perpendicular to $\Ab_0$. }
\label{fig:gauge}
\end{figure}

The integral term contributes to $A_\theta(r,\theta,\phi)$ when $B_\phi(r',\theta,\phi)$ is non-zero at some radius $r'$ between $r_0$ and $r$. Similarly, it contributes to $A_\phi(r,\theta,\phi)$ when $B_\theta(r',\theta,\phi)$ is non-zero. So the contribution to ${\cal A}$ from this term represents the net perpendicular magnetic flux beneath the field line concerned (Fig. \ref{fig:gauge}). For a field line with both footpoints on $r=r_0$, this is rather like choosing the curve $\gamma$ from Sect. \ref{sec:flux} to be the radial projection $\gamma'$ of the field line on $r=r_0$. Accordingly, this term will measure the twisting of magnetic field lines with height in the corona, and the net linking of flux beneath magnetic arcades. But it should be borne in mind that the projected curve $\gamma'$ will generally have $\int_{\gamma'}\Ab\cdot\eb_B\,\mathrm{d}l\neq 0$, and the additional contribution from the boundary term $r\Ab_0$ is needed to ensure that ${\cal A}$ gives an ideal-invariant flux.

With our choice of $\Ab_0$ in \eqref{eqn:A0}, the boundary term contribution to ${\cal A}$ represents the winding of coronal field lines around strong sources of magnetic flux on $r=r_0$. From \eqref{eqn:A0}, we see that the integral curves of $\Ab_0$ are the curves of constant $\psi$ \citep[cf.][]{hornig2006}. Since $\psi$ solves the Poisson equation with source term $B_r(r_0,\theta,\phi)$, these curves are analogous to the surfaces of equal temperature in a solution to the heat equation where $B_r(r_0,\theta,\phi)$ corresponds to a distribution of heat sources and sinks. Larger contributions to ${\cal A}$ arise when field lines (in projection) are aligned with these curves, which encircle the sources of (locally) strongest $|B_r|$.

So, overall, ${\cal A}$ in our gauge represents the net effect of two contributions: twisting of magnetic field lines with height, and winding around centres of strong flux on the boundary $r=r_0$. The examples of Sect. \ref{sec:nonaxi} suggest that both terms are generally significant.

\section{Magneto-frictional model} \label{sec:mf}

To approximate the evolution of non-potential magnetic fields in the corona, we use the magneto-frictional model introduced by \citet{vanballegooijen2000} and subsequently applied to the global corona by \citet{yeates2008}. In this model, the coronal magnetic field evolves through a continuous quasi-static sequence of approximately force-free equilibria, in response to continual shearing by photospheric footpoint motions. In this paper we use a uniform (but stretched) grid to cover the domain $\{r_0<r<r_1, \theta_0<\theta<\theta_1, 0<\phi<2\pi\}$ with a resolution $28\times 160\times 192$. Here we take $r_0 = R_\odot$ (the photosphere), $r_1=2.5R_\odot$, $\theta_0=0.05\pi$, and $\theta_1=0.95\pi$. Omitting the poles from our domain does not significantly affect the results presented here; the solution \eqref{eqn:psi} for $\psi$ remains valid provided that $\int_{\theta=\theta_0}^{\theta=\theta_1}\int_{\phi=0}^{\phi=2\pi} B_r \,\mathrm{d}\Omega = 0$, which we impose in our initial condition.

\begin{figure*}
\centering
\includegraphics[width=\textwidth]{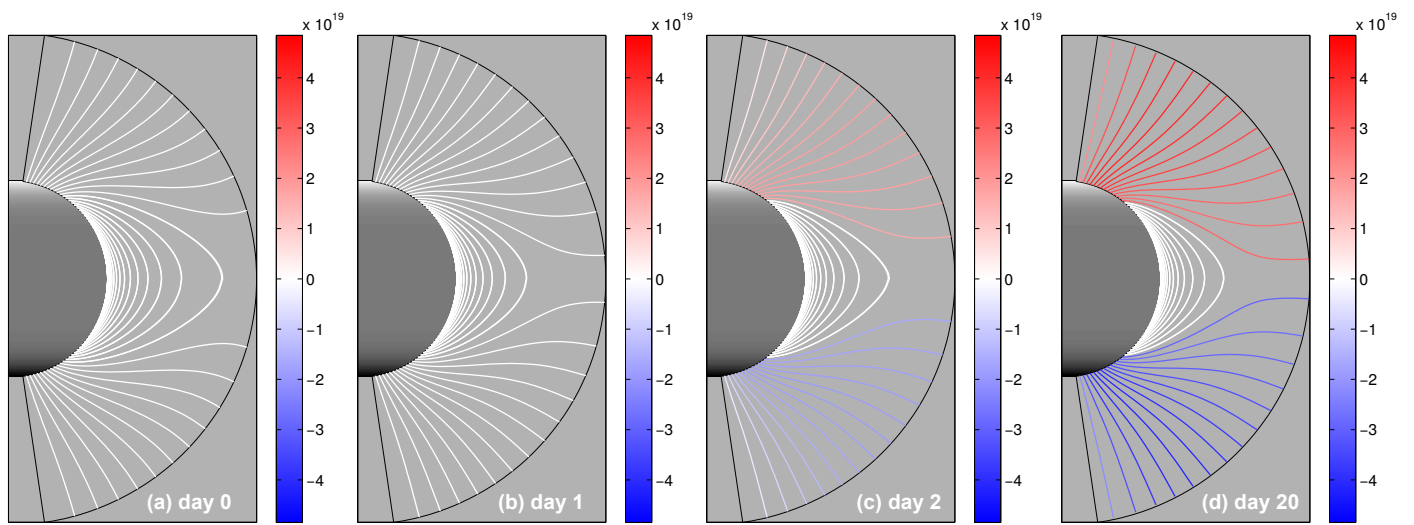}
\caption{Illustration of the dipolar simulation with $\nu_0=0.36\times 10^{-5}\,\mathrm{s}^{-1}$ on days 0, 1, 2, and 20. Greyscale shading on $r=r_0$ shows $B_r$ (white positive, black negative, saturated at $\pm 0.5\,\mathrm{G}$), and projected coronal magnetic field lines traced from height $r=R_\odot$ are coloured (red/blue) according to ${\cal A}$, saturated at $\pm 5\times 10^{19}\,\mathrm{Mx}$ with white indicating ${\cal A}\approx 0\,\mathrm{Mx}$.}
\label{fig:dipolefl}
\end{figure*}

\subsection{Coronal evolution}

In the magneto-frictional model, the coronal vector potential evolves according to the induction equation
\begin{equation}
\frac{\partial\Ab}{\partial t} = \vb\times\Bb - \eta\jb + \nabla\Phi,
\label{eqn:induc}
\end{equation}
where $\jb=\nabla\times\Bb$ and $\eta$ represents a turbulent resistivity arising from the cumulative effect of small-scale coronal flows. For simplicity in this paper, we follow \citet{mackay2006} and set
\begin{equation}
\eta = \eta_0\left(1 + 0.2\frac{|\jb|}{\max|\Bb|}\right),
\end{equation}
where $\eta_0$ is a constant background value and the second term acts only in regions of strong current density $|\jb|$ to limit the formation of unresolved gradients in $\Bb$. An alternative would be to consider higher-order hyperdiffusion \citep[as in][]{yeates2014}, but the simpler form suffices here. 

The gauge $\Phi$ in \eqref{eqn:induc} is, of course, arbitrary. For computation itself we use the Weyl gauge $\Phi\equiv 0$, but for calculating the field line helicity we  subsequently recompute the DeVore gauge $\Ab$ from $\Bb$, as defined in Section \ref{sec:gauge}.

The main simplification in the magneto-frictional method is to forego solving the full MHD equations and instead approximate the plasma velocity by
\begin{equation}
\vb = \nu\frac{\jb\times\Bb}{|\Bb|^2} + v_{\rm out}\left(\frac{r}{r_1}\right)^{11.5}\eb_r.
\label{eqn:v}
\end{equation}
Here the first term is a friction-like term that enforces relaxation towards a force-free equilibrium. The factor $|\Bb|^2$ prevents relaxation from being inhibited in weak-field regions, although it must be limited away from zero near null points where $|\Bb|=0$. The coefficient $\nu$ has the same dimensions as $\eta$, and is set to \mbox{$\nu = \nu_0 r^2\sin^2\theta$}. The second term in \eqref{eqn:v} is a radial outflow imposed only near the outer boundary. This term simulates (crudely) the effect of the solar wind in radially opening out the magnetic field lines, while allowing horizontal field to pass through the upper boundary if necessary.

Equations \eqref{eqn:induc} and \eqref{eqn:v} are solved on a staggered grid \citep{yee1966} using finite differences. Zero-gradient boundary conditions are imposed at $r=r_1$, and $B_\theta=0$ is imposed at $\theta=\theta_0, \theta_1$. At $r=r_0$, we do not prescribe $\vb$ according to Eqn. \eqref{eqn:v}, but rather determine $\partial A_\theta/\partial t$ and $\partial A_\phi/\partial t$ from our imposed photospheric driving. (No boundary condition on $A_r$ is needed, owing to the staggered grid.) For a given photospheric driver, the coronal model is then determined by three parameters: $\nu_0$, $\eta_0$ and $v_{\rm out}$. The friction coefficient $\nu_0$ controls the speed of coronal relaxation relative to the surface evolution, while $\eta_0$ controls the rate of coronal diffusion. Rather than $\eta_0$, we vary the dimensionless number $\eta_0/(R_\odot^2\nu_0)$, which measures the relative importance of diffusion compared to relaxation in the corona \citep[cf.][]{cheung2012}. For this paper, we fix the radial outflow speed $v_{\rm out}=100\,\mathrm{km}\,\mathrm{s}^{-1}$.

\subsection{Photospheric driving}

The magneto-frictional method simulates the evolution of the coronal magnetic field in response to shearing by surface motions. In this paper, we consider the effect of these motions on three different initial magnetic fields. The motions are modelled by a simple surface flux transport model \citep{sheeley2005,mackay2012,jiang2014} in which, at $r=r_0$, we impose
\begin{align}
\frac{\partial A_\theta}{\partial t} &= r\sin\theta\,\Omega(\theta)B_r - \frac{D}{r_0\sin\theta}\frac{\partial B_r}{\partial\phi},\label{eqn:atheta}\\
\frac{\partial A_\phi}{\partial t} &= - \frac{D}{r_0}\frac{\partial B_r}{\partial\theta}.
\end{align}
The first term in \eqref{eqn:atheta} represents differential rotation. The simulations are carried out in the carrington frame, and we choose the \citet{snodgrass1983} angular velocity (in degrees per day)
\begin{equation}
\Omega(\theta) = 0.18 - 2.3\cos^2\theta - 1.62\cos^4\theta.
\end{equation}
This implies that the coronal magnetic field is relaxing relative to the carrington frame, rather than to the background stars. The coefficient $D=600\,\textrm{km}\,\textrm{s}^{-1}$ represents ``supergranular diffusion'' of $B_r$, namely the net large-scale effect of the random walk of magnetic elements under supergranular convection on the solar surface. For illustrative purposes, we neglect other flux transport effects such as meridional flow, as well as the emergence of new magnetic flux.

\section{Dipolar field} \label{sec:dipolar}

\begin{figure}
\resizebox{\hsize}{!}{\includegraphics{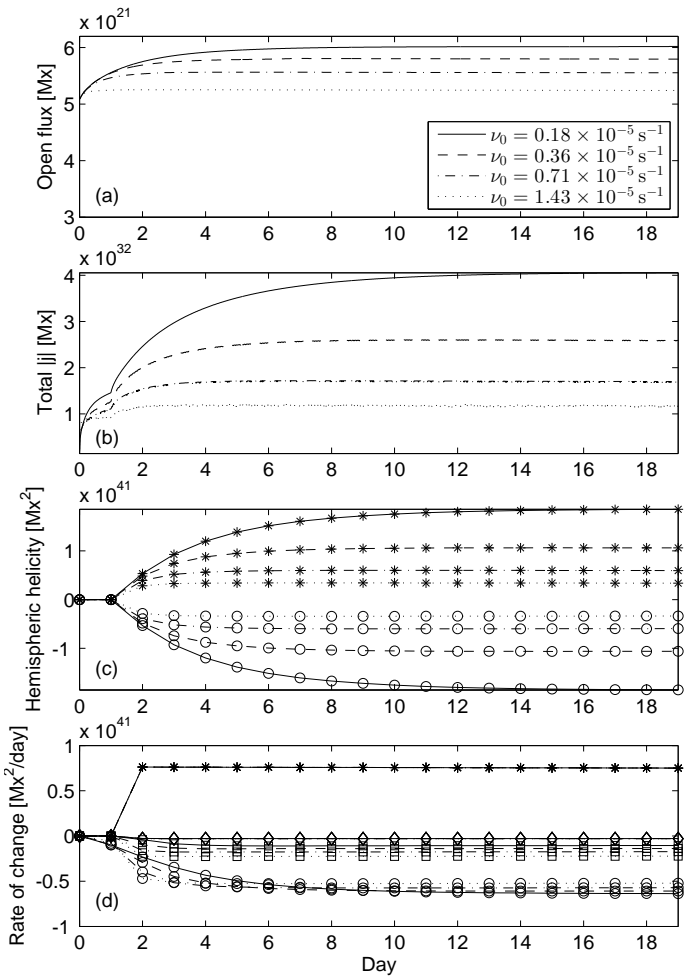}}
\caption{Various integrated quantities as a function of time, for the dipolar simulations with different $\nu_0$ (indicated by line styles). Panel (a) shows the open magnetic flux $\int_{r=r_1}|B_r|\,\mathrm{d}\Omega$, panel (b) shows $\int_D|\jb|\,\mathrm{d}V$, and panel (c) shows $H_{\rm N}$ (asterisks) and $H_{\rm S}$ (circles). Panel (d) shows the terms in equation \eqref{eqn:dhdt} for the northern hemisphere, with asterisks denoting $S_0$, circles $S_1$, squares $S_{\rm eq}$, and diamonds $S_V$.}
\label{fig:dipolediags}
\end{figure}

\begin{figure}
\resizebox{\hsize}{!}{\includegraphics{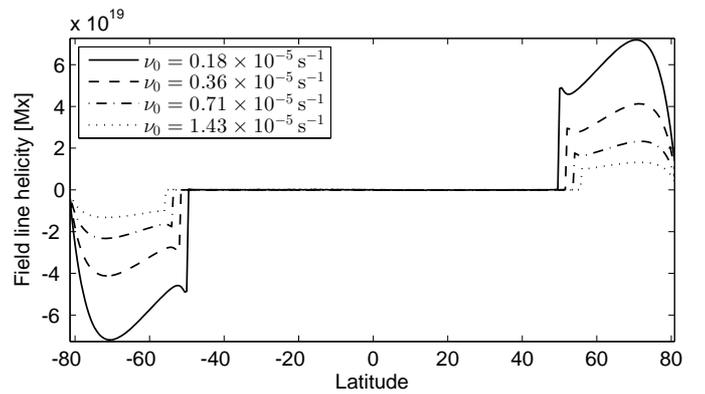}}
\caption{Latitudinal distribution of field line helicity for the dipolar simulations, showing how the peak value tends to zero as the friction parameter $\nu_0$ is successively doubled. A log-log fit shows that ${\cal A}_{\rm max}\sim \nu_0^{-0.8}$.}
\label{fig:dipoleflh}
\end{figure}

Our first simulation is initialized with a potential field extrapolation from the photospheric boundary condition
\begin{equation}
B_r(r_0,\theta,\phi) = B_0\cos^7\theta,
\end{equation}
where $B_0=1\,\mathrm{G}$, along with $B_\theta=0$ on the boundaries $\theta=\theta_0,\theta_1$ and $B_\theta=B_\phi=0$ on the outer boundary $r=r_1$. The potential field is computed using the eigenfunction method of \citet{vanballegooijen2000}. The coronal magnetic field is then evolved with magneto-friction, as described in Sect. \ref{sec:mf}. For this example the coronal field remains close to potential, with low electric currents, so the results are insensitive to $\eta_0$. Accordingly we will illustrate only the effect of varying $\nu_0$, while holding the dimensionless ratio $\eta_0/(R_\odot^2\nu_0)$ fixed at $2.89\times 10^{-5}$ (a typical value from previous simulations). For the first day of evolution, no photospheric motions are applied, so as to illustrate the effect of switching on differential rotation from day 1 onwards.

The evolution of the magnetic field structure for one of the runs is shown in Fig. \ref{fig:dipolefl}. The evolution is straightforward: firstly there is an opening out of the magnetic field, due to the radial outflow at the upper boundary. This expansion takes approximately 1 day, and creates electric currents near the outer boundary associated with the extended ``streamer'' structure at the equator. Once the surface motions are switched on, the field then relaxes to a dynamical equilibrium between the footpoint shearing and the magneto-frictional relaxation. As we will discuss below, there is non-zero field line helicity associated with this dynamical equilibrium (shown by the colours in Fig. \ref{fig:dipolefl}). The time taken to reach equilibrium depends on $\nu_0$.

Indeed, it is instructive to consider the effect of $\nu_0$ (the rate of frictional relaxation) on the evolution. Fig. \ref{fig:dipolediags} shows a number of integrated quantities, as a function of time for four runs with different $\nu_0$. As seen in Fig. \ref{fig:dipolediags}(a), weaker friction allows the radial outflow to open out the field further, leading to more open magnetic field lines, although this is not a particularly strong effect. More striking, in this example, is the increased coronal electric current that weaker friction allows (Fig. \ref{fig:dipolediags}b). A small part of this difference in current arises from the greater initial expansion of the field, but most arises from the character of the dynamical equilibrium. When friction is weaker, the field lines relax back less in response to shearing of their footpoints by differential rotation, so more current is stored in the corona. The shearing of field lines in the equilibrium is actually rather small and hard to discern in Fig. \ref{fig:dipolefl}, although it is just visible at the south pole on day 20.

Next we consider the evolution of magnetic helicity. In this unusually symmetric situation, it is helpful to consider the net helicity in each hemisphere, 
\begin{equation}
H_{\rm N} = \int_{\theta < \pi/2}\Ab\cdot\Bb\,\mathrm{d}V, \qquad H_{\rm S} = \int_{\theta > \pi/2}\Ab\cdot\Bb\,\mathrm{d}V.
\end{equation}
By symmetry these are equal and opposite (so that the total helicity vanishes). They are shown in Fig. \ref{fig:dipolediags}(c). Before the surface motions are switched on there is no helicity, since $B_\phi\equiv 0$ and $A_\theta\equiv 0$. After the motions are switched on, the helicity increases to a steady value in each hemisphere. It is clear from Fig. \ref{fig:dipolediags}(c) that this steady value is larger when the friction is weaker, in accordance with the greater shear of the equilibrium field lines.

The equilibrium distribution of field line helicity ${\cal A}$ is shown both by the colours in Fig. \ref{fig:dipolefl} and, as a function of latitude, in Fig. \ref{fig:dipoleflh}. It is clear that the helicity in each hemisphere is not distributed uniformly among all field lines, but is stored only on open field lines. This is due to the symmetry of the configuration: closed field lines cross the equator and pick up equal and opposite contributions to ${\cal A}$ from each hemisphere.

\begin{figure}
\resizebox{\hsize}{!}{\includegraphics{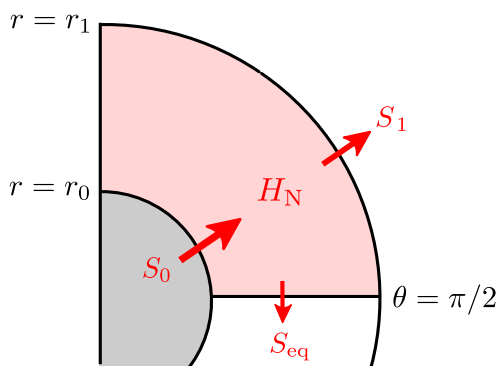}}
\caption{Schematic of helicity flow in the dipolar example, for the northern hemisphere. Arrows show the direction of positive helicity transfer as measured by the surface terms $S_0$, $S_1$, and $S_{\rm eq}$ in \eqref{eqn:dhdt}.}
\label{fig:hemi}
\end{figure}

\begin{figure*}
\centering
\includegraphics[width=\textwidth]{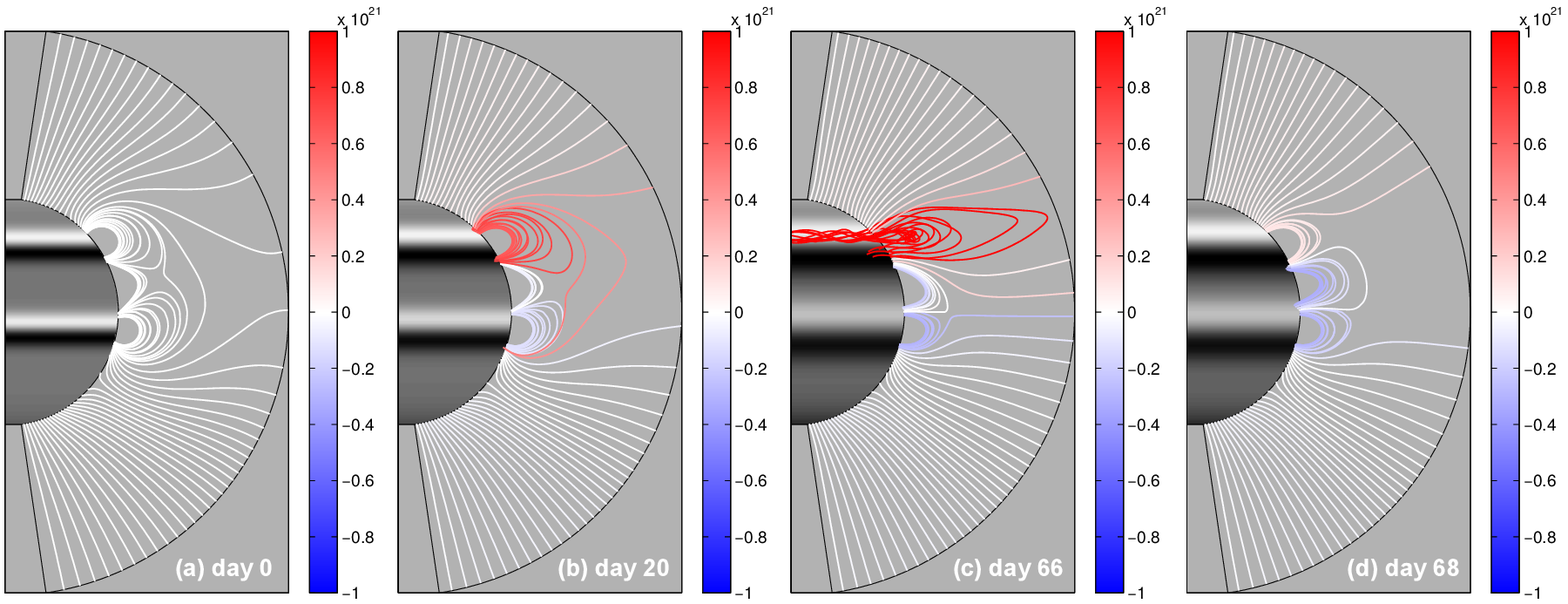}
\caption{Illustration of the quadrupolar simulation with $\nu_0=0.36\times 10^{-5}\,\mathrm{s}^{-1}$ on days 0, 20, 66, and 68. Greyscale shading on $r=r_0$ shows $B_r$ (white positive, black negative, saturated at $\pm 2\,\mathrm{G}$), and projected coronal magnetic field lines traced from height $r=1.2R_\odot$ are coloured (red/blue) according to ${\cal A}$, saturated at $\pm 10^{21}\mathrm{Mx}$ with white indicating ${\cal A}\approx 0\,\mathrm{Mx}$.}
\label{fig:quadrupolefl}
\end{figure*}

It is also interesting to consider the helicity flux through the boundaries. To calculate this, it is most convenient to use the form
\begin{equation}
\frac{dH}{dt} = -2\int_V\Eb\cdot\Bb\,\mathrm{d}V + \oint_{\partial V}\Ab\times\left(2\Eb + \frac{\partial\Ab}{\partial t}\right)\cdot\,\mathrm{d}{\boldsymbol a},
\label{eqn:dhdt}
\end{equation}
which is easily derived using Faraday's law
\begin{equation}
\frac{\partial \Bb}{\partial t} = -\nabla\times\Eb.
\end{equation}
This is valid for the helicity in any subdomain $V$, whether magnetically closed or not. Here we use $\Ab$ computed in our gauge to estimate $\partial\Ab/\partial t$, and we also record the electric field $\Eb$ during the simulation. When we apply this formula to $H_{\rm N}$ (or $H_{\rm S}$), we obtain four contributions: the volume dissipation term $S_V$, and three contributions to the surface integral from different boundaries, namely $S_0$ ($r=r_0$), $S_1$ ($r=r_1$), and $S_{\rm eq}$ ($\theta=\pi/2$), as in Fig. \ref{fig:hemi}. There is no contribution from the latitudinal boundaries $\theta=\theta_0$ and $\theta=\theta_1$ owing to our boundary conditions in the simulation. Fig. \ref{fig:dipolediags}(d) shows these four contributions for each of the dipole simulations, for the northern hemisphere. (The southern hemisphere contributions are equal and opposite.) Firstly, the volume dissipation term is small compared to the surface terms. The main contributions are an injection $S_0$ of helicity through $r=r_0$, by differential rotation, and an output $S_1$ through the upper boundary. The latter would correspond to winding up of the solar wind (the Parker spiral). However, the helicity output is rather less than the input ($6.1\times 10^{40}\,\mathrm{Mx}^2\,\mathrm{day}^{-1}$ compared to $7.5\times 10^{40}\,\mathrm{Mx}^2\,\mathrm{day}^{-1}$ for $\nu_0=0.36\times 10^{-5}\,\mathrm{s}^{-1}$). The difference is accounted for by $S_{\rm eq}$, which represents a net transfer of helicity across the equator on closed field lines. During the relaxation phase, there is a slight imbalance between these terms, allowing the equilibrium helicity to build up in each hemisphere. The overall flow of helicity is summarised in Fig. \ref{fig:hemi}.

Note that, as the friction parameter $\nu_0$ is increased, the stored field line helicity in each hemisphere, along with $H_{\rm N}$ and $H_{\rm S}$, tends to zero approximately as $\nu_0^{-0.8}$. However, helicity is injected by differential rotation through the photosphere at a constant rate $S_0$ that is independent of $\nu_0$. Fig. \ref{fig:dipolediags}(d) shows that the lack of stored helicity is compensated by the other surface terms $S_1$ and $S_{\rm eq}$ during the relaxation phase. Even for the finite values of $\nu_0$ considered here, the stored helicity in the corona is little more than the helicity injected in a single day by differential rotation. However, we will see in the subsequent examples that much more helicity can be stored if we break the symmetry of the magnetic configuration.

It is interesting to note that the sign of helicity injected into the solar wind is opposite to that of \citet{berger2000}, who estimated the injection of helicity into the volume $r>r_0$ by solar rotation. Figure \ref{fig:hemi} shows that the outward helicity flux in the northern hemisphere is positive in our example, since $S_1$ is negative (Fig. \ref{fig:dipolediags}d). This sign is opposite to the direction of winding of the Parker spiral. However, this is an apparent difference caused by our use of the carrington frame. If the constant 27-day rotation rate were added back in, the sign would reverse.

\section{Quadrupolar field} \label{sec:quadrupolar}

In more realistic configurations, differential rotation is able to build up field line helicity on closed field lines. Our second axisymmetric example gives a simple demonstration of this process, starting from a potential field extrapolated from the photospheric distribution
\begin{align}
B_r(\theta,\phi) &= B_0\cos^7\theta + B_1\big(\cos\theta - \cos\theta_1\big)\exp\left[-\frac{(\cos\theta - \cos\theta_1)^2}{d^2}\right] \nonumber\\
&+ B_2\big(\cos\theta - \cos\theta_2\big)\exp\left[-\frac{(\cos\theta - \cos\theta_2)^2}{d^2}\right].
\end{align}
With $B_0=1\,\mathrm{G}$, $B_1=B_2=100\,\mathrm{G}$, the bipolar rings each contain the same unsigned magnetic flux $2\pi d^2R_\odot^2$, provided that they overlap neither each other nor the poles. We take $d=0.1$, and locate them at $\theta_1 = 0.3\pi$ and $\theta_2=0.55\pi$ (illustrated in Fig. \ref{fig:quadrupolefl}). Since the rings are asymmetrically placed with respect to the equator, we expect differential rotation above each PIL to build up helicity at different rates, even though both rings contain the same magnetic flux. The same photospheric motions are imposed as in Sect. \ref{sec:dipolar}, except that they are switched on immediately. A slightly different value $\eta_0/(R_\odot^2\nu_0)=3.47\times 10^{-5}$ is used, although we will also consider the effect of varying this parameter below.

\begin{figure}
\resizebox{\hsize}{!}{\includegraphics{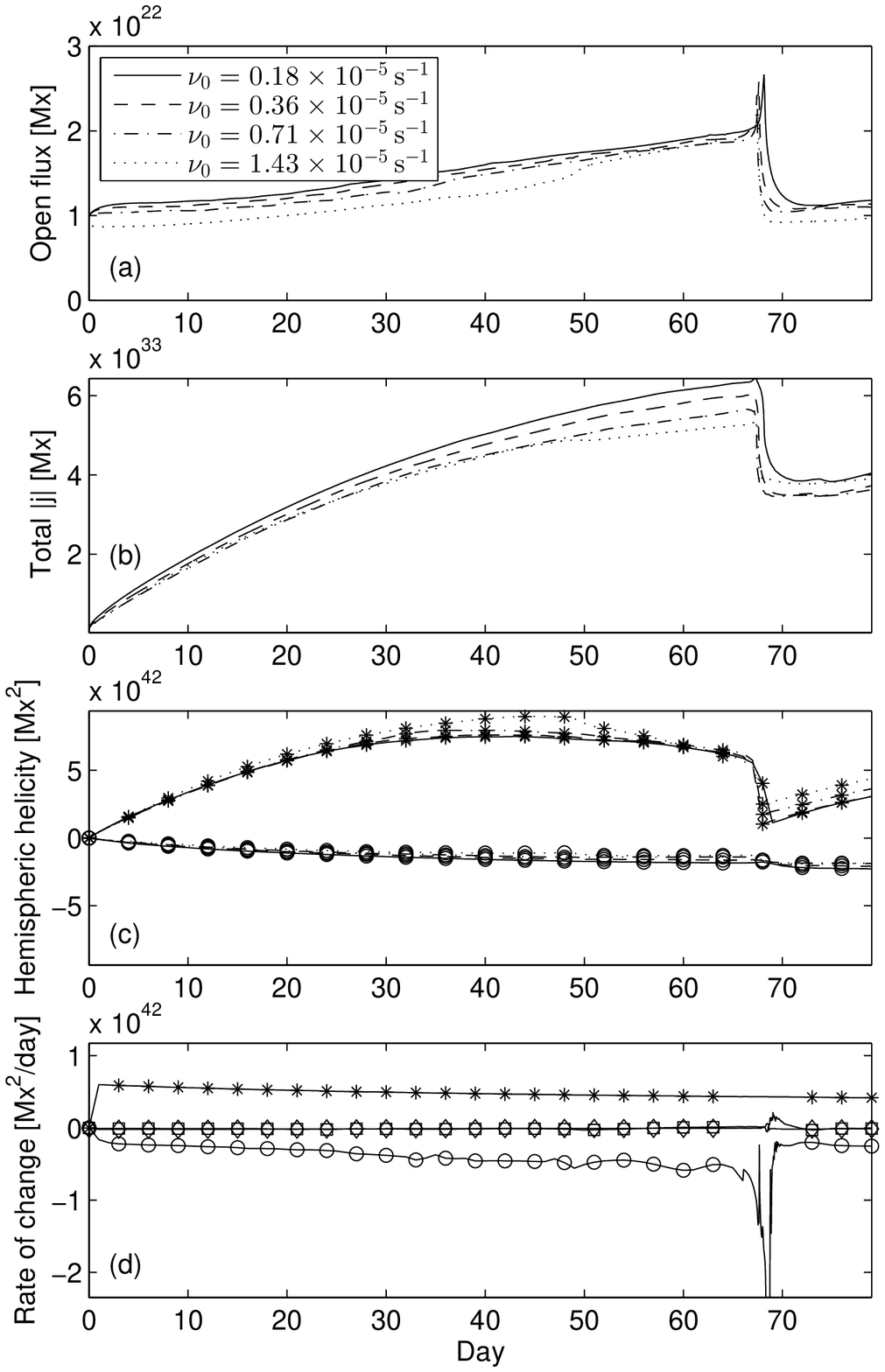}}
\caption{Various integrated quantities as a function of time, for the quadrupolar simulations with different $\nu_0$ (indicated by line styles). The format is the same as Figure \ref{fig:dipolediags}. For clarity, panel (d) shows only the run with $\nu_0=0.36\times 10^{-5}\,\mathrm{s}^{-1}$, and only for the northern hemisphere, although the hemispheres are no longer symmetric. The peak value of $S_1$ during the flux rope eruption is not shown, and is much larger, about $-2.7\times 10^{43}\,\mathrm{Mx}^2\,\mathrm{day}^{-1}$.}
\label{fig:quadrupolediags}
\end{figure}

Figure \ref{fig:quadrupolefl} shows how the magnetic field evolves over 68 days, while Fig. \ref{fig:quadrupolediags} shows various integrated quantities, analogous to Fig. \ref{fig:dipolediags}. The most striking difference from the dipolar case is that the quadrupolar system does not reach a dynamical equilibrium, in spite of the fact that the rate $S_0$ of helicity injection by differential rotation remains fairly constant, albeit higher than before owing to the greater magnetic flux on $r=r_0$. (The slight decay in $S_0$ over time arises from diffusive decay of the more concentrated photospheric field, visible in Fig. \ref{fig:quadrupolefl}.) Instead, current and helicity continue to be injected into the corona. The open flux does initially level off, but then increases as the magnetic arcades are sheared and energised.

It is clear from Fig. \ref{fig:quadrupolefl} that the additional helicity is primarily stored along closed field lines, particularly those that do not cross the equator. This arises because the footpoints are no longer symmetric about the equator, so that differential rotation shears the magnetic loops. The sign injected is opposite for the arcades in each hemisphere. More helicity is injected in the northern hemisphere, simply because the two bipolar rings are asymmetrically placed and the northern ring lies at a latitude with greater shear in the differential rotation. This asymmetry also leads, eventually, to negative helicity in the arcade straddling the equator (Fig. \ref{fig:quadrupolefl}). For finite $\nu_0$, the open field lines near the poles actually store a similar amount of field line helicity as in the dipolar example, but this is insignificant compared to that stored at lower latitudes. Moreover, this lower latitude helicity is almost independent of $\nu_0$, since it is enforced topologically by the footpoint motions and cannot be removed by ideal relaxation, however rapid. As the open flux increases gradually with energisation of the field, the helicity output $S_1$ increases, consistent with the dipolar example where open field lines act as continuous ``conduits'' of helicity from the photosphere out to the solar wind. The increasing proportion of open field lines leads to the levelling off of hemispheric helicity from about day 30 onwards. In this example, the cross-equatorial helicity flux $S_{\rm eq}$ is negligible.

We remark that the sign of helicity in each hemisphere, in this example, is opposite to the typical hemispheric pattern of helicity on the Sun, which is negative in the northern hemisphere and positive in the south \citep{pevtsov2003}. This arises from the East-West orientation of the polarity inversion line in our axisymmetric model. On the real Sun, polarity inversion lines at active latitudes are often aligned North-South, so that differential rotation injects helicity of the observed majority sign. This was illustrated by the simulations of \citet{devore2000} in \edit{Cartesian} geometry, and  \citet{yeates2009a} in spherical geometry.

As is evident in Fig. \ref{fig:quadrupolediags}, the amount of electric current and helicity in the corona does not build up indefinitely, but is suddenly reduced on about day 67 of the simulation. This sudden reduction results from ejection of the magnetic flux rope that forms above the northern polarity inversion line. The flux rope is visible on day 66 in Fig. \ref{fig:quadrupolefl}, but has been ejected through the outer boundary $r=r_1$ by day 68, leaving only a weakly sheared arcade behind it. The mechanism by which the flux rope forms is well understood \citep{vanBallegooijen1989}; essentially, it is a combination of reconnection of sheared magnetic loops accompanied by flux cancellation due to supergranular diffusion on $r=r_0$, which leaves horizontal magnetic field in the corona above polarity inversion lines. Due to the symmetry in this rather artificial example, the flux rope that forms is detached from the photosphere, encircling the whole Sun. Since the corresponding magnetic field lines are either closed or ergodic (infinite length), the field line helicity in the rope is undefined. (In Fig. \ref{fig:quadrupolefl}, the colour scale is saturated at $\pm 10^{21}\,\mathrm{Mx}$.) Nevertheless, integrating for a finite length clearly indicates the location of the rope. As is evident in Fig. \ref{fig:quadrupolediags}(d), the eruption causes a very high, sudden, spike in the helicity output $S_1$, and a consequent sudden reduction in the total helicity $H_{\rm N}$ in the northern hemisphere (Fig. \ref{fig:quadrupolediags}c). After the eruption, the helicity begins to build up again since the footpoint shearing continues.

Finally we consider the effect of the simulation parameters $\nu_0$ and $\eta_0$. We have already seen that the hemispheric helicity and open flux depend only weakly on $\nu_0$.  We have also run simulations with different $\eta_0/(R_\odot^2\nu_0)$ but the same value of $\nu_0=0.36\times 10^{-5}\,\mathrm{s}^{-1}$. We find that increasing $\eta_0/(R_\odot^2\nu_0)$ from $1.73\times 10^{-5}$ to $6.94\times 10^{-5}$ delays the flux rope eruption by three days, but has little impact on the magnitude of open flux or hemispheric helicity overall. This is consistent with the findings of \citet{yeates2009}, who showed that higher diffusion limits the speed at which highly twisted flux ropes are able to form, by dissipating the concentrated electric currents in the ropes. This led to a lower eruption rate.

This example has shown how field line helicity reveals the storage of helicity on closed magnetic field lines in the corona, as well as the sudden expulsion of this helicity in the form of flux rope eruptions. In the next section, we see these processes at work in a more realistic global configuration.

\section{Non-axisymmetric field} \label{sec:nonaxi}

\begin{figure}
\resizebox{\hsize}{!}{\includegraphics{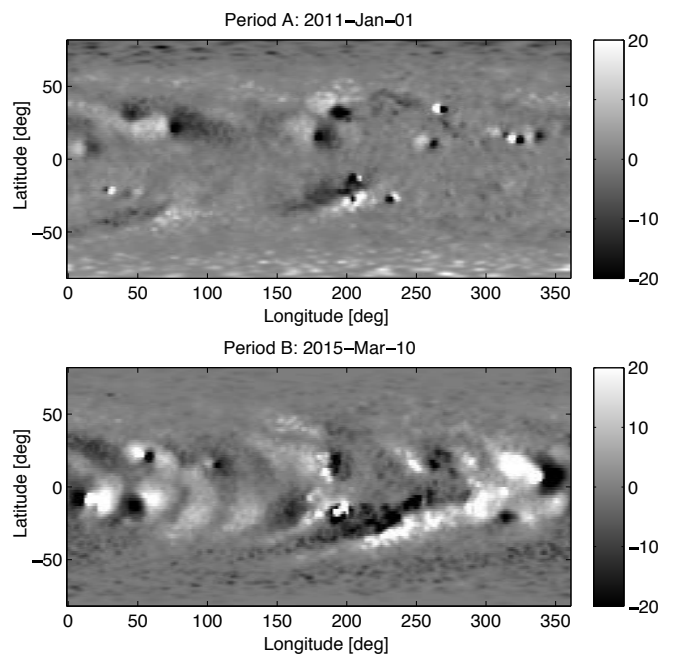}}
\caption{ADAPT maps of $B_r$ on $r=r_1$, used to generate the initial potential field extrapolations for periods A and B (white positive, black negative, saturated at $\pm 20\,\mathrm{G}$).}
\label{fig:nonaxibr}
\end{figure}

Our final example is a more realistic global magnetic configuration. For simplicity, we still consider only driving by large-scale surface motions, and continue to neglect the emergence of new magnetic flux.

Two simulations are presented: period A and period B. Each starts from a potential field extrapolation, as before, but now these are computed from full-surface $B_r$ maps modelling the real Sun on two dates: 2011-Jan-01 and 2015-Mar-10 (Fig. \ref{fig:nonaxibr}). The maps are taken from the Air Force Data-Assimilative Photospheric Flux Transport (ADAPT) model \citep{arge2010,henney2012,hickmann2015}, which assimilates observed magnetograms for the visible side of the Sun into a surface flux transport model. Here we simply take the two ADAPT maps as initial conditions for our two simulations. The maps in question come from ADAPT runs based on GONG magnetograms, and have been remapped to our simulated grid. A multiplicative flux correction has been applied to ensure flux balance. The choice of ADAPT maps, as opposed to any other model, is not particularly important; we simply wanted a realistic distribution of magnetic flux on the full solar surface. Period B represents a more active time in the solar cycle than Period A, with considerably larger total flux. By this time in early 2015, the polar fields visible in Period A have been almost completely removed by cancellation with magnetic flux from Cycle 24 active regions.

The magneto-frictional simulations use the same parameters as Sect. \ref{sec:quadrupolar}, driven by the same differential rotation and supergranular diffusion, with the parameters $\nu_0=0.36\times 10^{-5}\,\mathrm{s}^{-1}$ and $\eta_0/(R_\odot^2\nu_0)=3.47\times 10^{-5}$. The evolution is followed for much longer, up to 180 days. Whilst it is unrealistic to evolve the global magnetic field for so long without any new flux emergence, our purpose is to explore how the field line helicity responds  to photospheric motions. On the Sun, the large-scale magnetic fields at higher latitudes do indeed result from many months of evolution of old active region fields \citep{petrie2015}. 

\begin{figure*}
\centering
\includegraphics[width=0.45\textwidth]{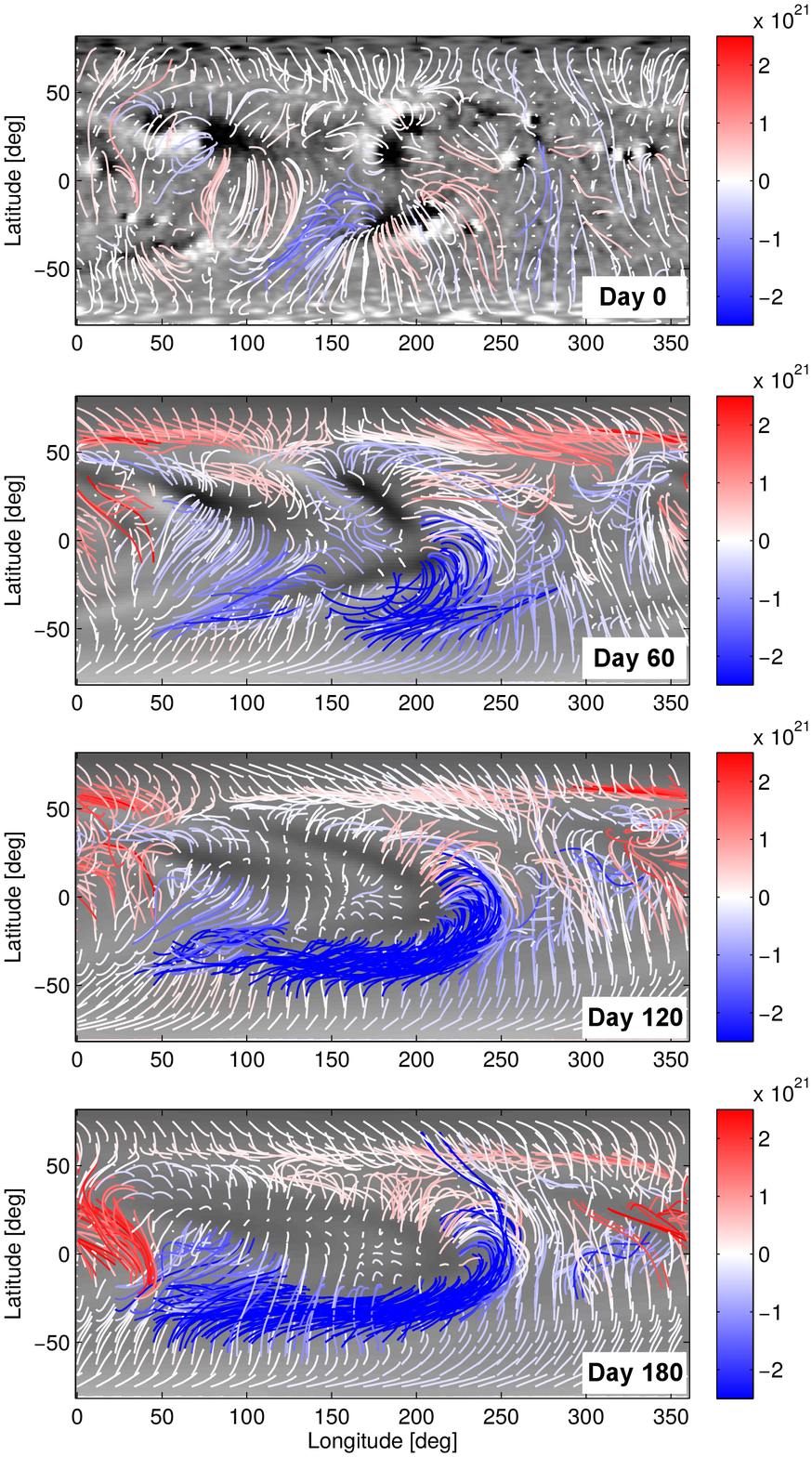}
\includegraphics[width=0.45\textwidth]{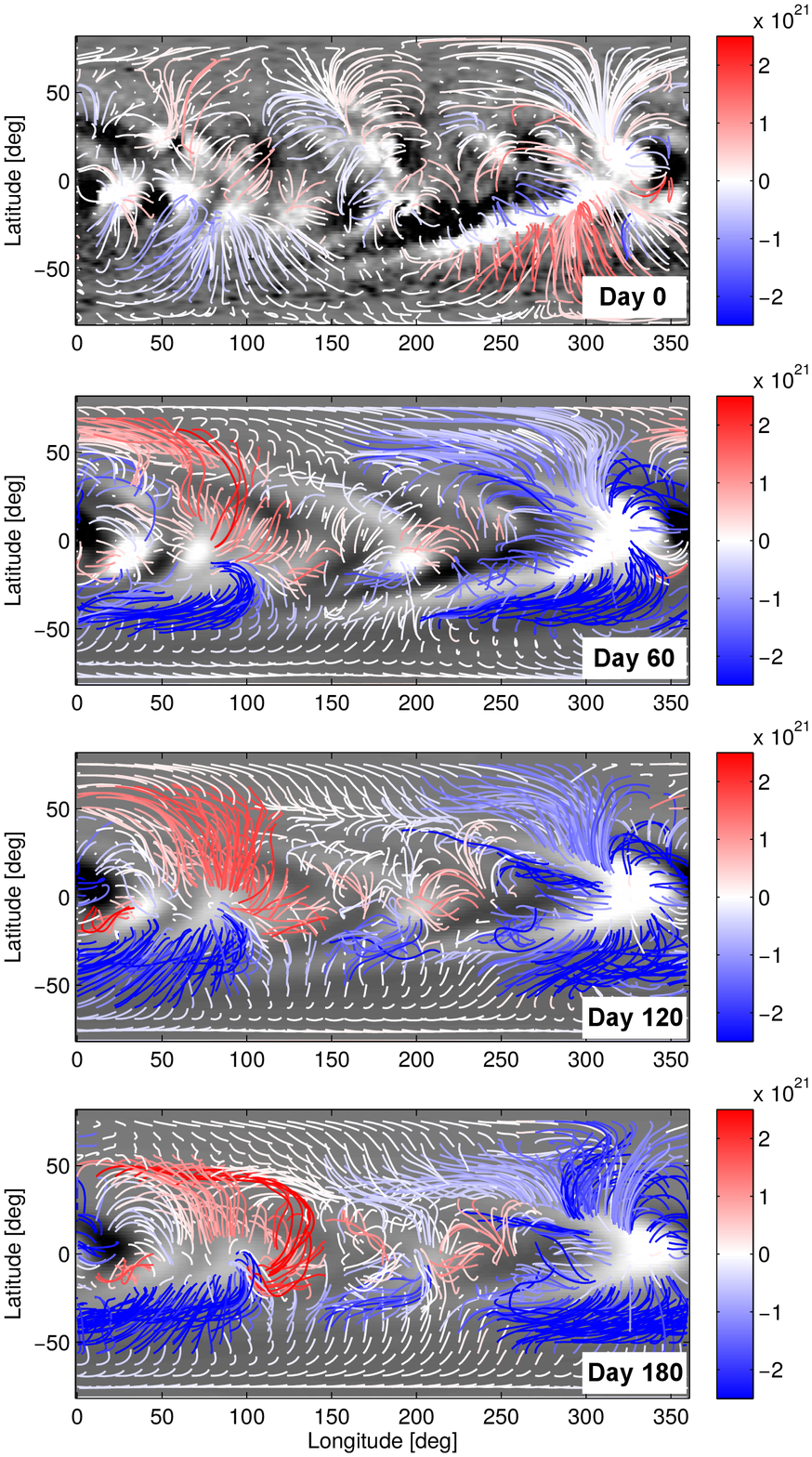}
\caption{Projected magnetic field lines in the period A (left column) and B (right column) simulations, on days 0, 60, 120, and 180. Greyscale shading on $r=r_0$ shows $B_r$ (white positive, black negative, saturated at $\pm 10\,\mathrm{G}$), and projected coronal magnetic field lines traced from height $r=r_0$ are coloured (red/blue) according to ${\cal A}$, saturated at $\pm 2.5\times 10^{21}\,\mathrm{Mx}$. \textbf{Animated versions of these sequences are available in the online journal.}}
\label{fig:nonaxifl2d}
\end{figure*}

Figure \ref{fig:nonaxifl2d} shows four snapshots of the magnetic field during each period, along with the field line helicity on a grid of magnetic field lines. The effects of both differential rotation and of supergranular diffusion are apparent on $r=r_0$, where the pattern of $B_r$ is both sheared and significantly smoothed out, removing smaller features. Once again, we see how field line helicity is injected and stored in closed magnetic arcades. However, the distribution of ${\cal A}$ among closed field lines is far from uniform. The amount of field line helicity stored in any particular magnetic arcade is dependent on the degree of shearing of the arcade, which depends both on the orientation and the $B_r$ pattern on $r=r_0$ \citep[see also][]{yeates2009a}. In fact, the non-uniform distribution of magnetic flux across the solar surface means that some field lines have non-zero ${\cal A}$ even in the initial potential field (day 0 for each period in Fig. \ref{fig:nonaxifl2d}). But much stronger field line helicity builds up at particular locations where the field orientation is favourable to shearing by differential rotation. This often reverses the initial sign of ${\cal A}$ at a particular location (e.g., around $300^\circ$ longitude in the southern hemisphere between days 0 and 60 of period B). As in the quadrupolar example (Sect. \ref{sec:quadrupolar}), the largest values of $|{\cal A}|$ lie in twisted magnetic flux ropes.

\begin{figure}
\resizebox{\hsize}{!}{\includegraphics{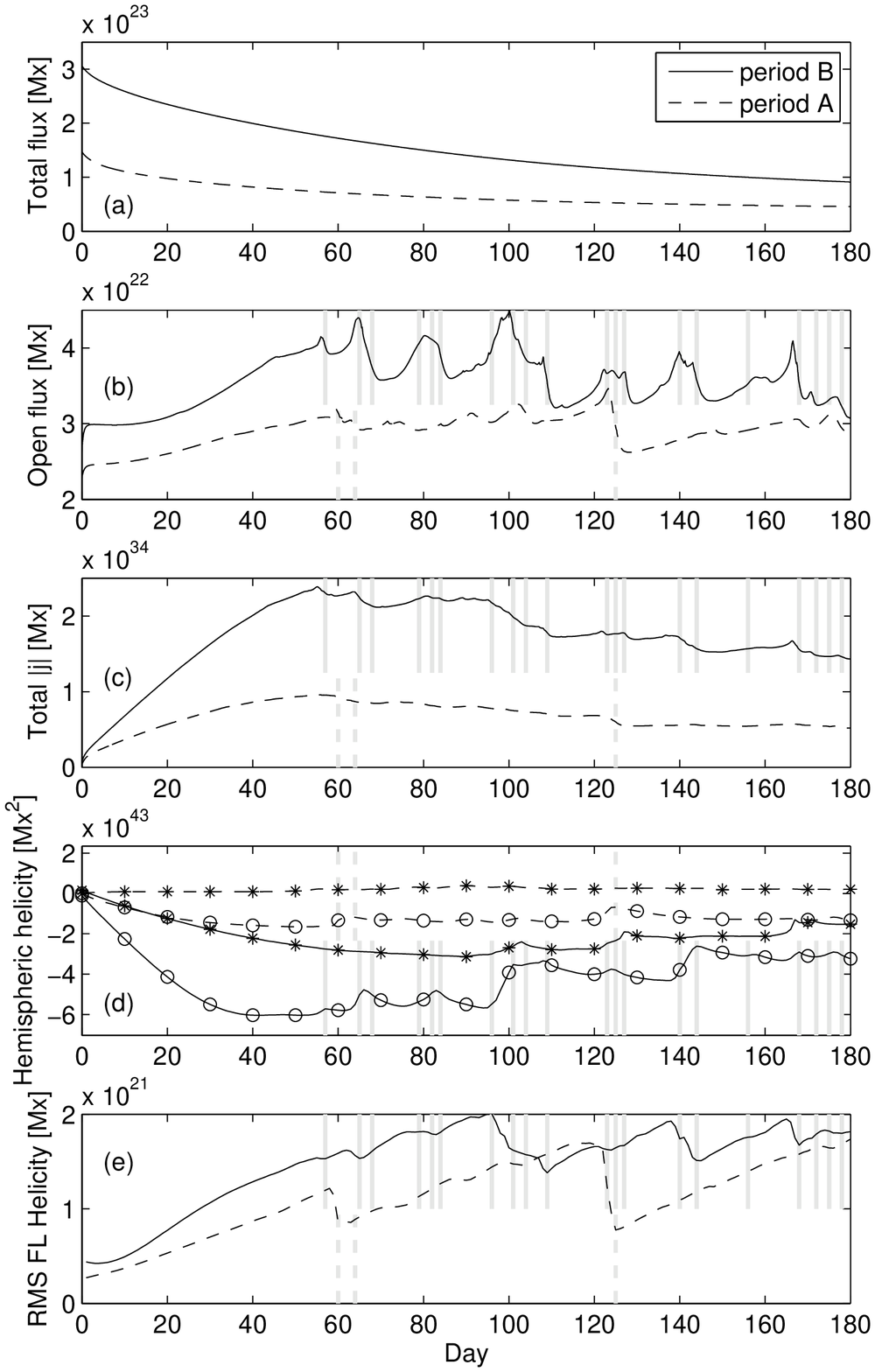}}
\caption{Various integrated quantities as a function of time, for the non-axisymmetric simulations (periods A and B). Panel (a) shows the total photospheric magnetic flux $\int_{r=r_0}|B_r|\,\mathrm{d}\Omega$, panel (b) shows the total open flux $\int_{r=r_1}|B_r|\,\mathrm{d}\Omega$, panel (c) shows $\int_D|\jb|\,\mathrm{d}V$, panel (d) shows $H_{\rm N}$ (asterisks) and $H_{\rm S}$ (circles), and panel (e) shows the root-mean-square field line helicity $(\int_D{\cal A}^2\,\mathrm{d}V/\int_D\,\mathrm{d}V)^{1/2}$. The vertical grey lines indicate times of strong flux rope ejections, as explained in the text.}
\label{fig:nonaxidiags}
\end{figure}

Figure \ref{fig:nonaxidiags} shows how global quantities evolve in the simulations for periods A and B. In the initial map there is about twice as much flux in period B as in period A, leading to correspondingly higher open flux, total current, and hemispheric helicity throughout the simulation. The total current and open flux increase gradually over about the first 60 days as the coronal field is energised, before reaching saturation and then gradually decaying over the rest of the simulation (owing to the decaying photospheric flux). The root-mean-square ${\cal A}$ also takes about 2 months to reach its maximum value, although this does not seem to decay over the remainder of the simulation. This indicates how helicity is stored in the coronal magnetic field through memory of the footpoint motions. The hemispheric helicities are harder to interpret, highlighting the greater utility of ${\cal A}$ as a diagnostic when helicity is non-uniformly distributed through the corona. Nevertheless, it is generally true that the helicity has greater magnitude in period B than in period A, commensurate with the greater flux.

\begin{figure*}
\centering
\includegraphics[width=\textwidth]{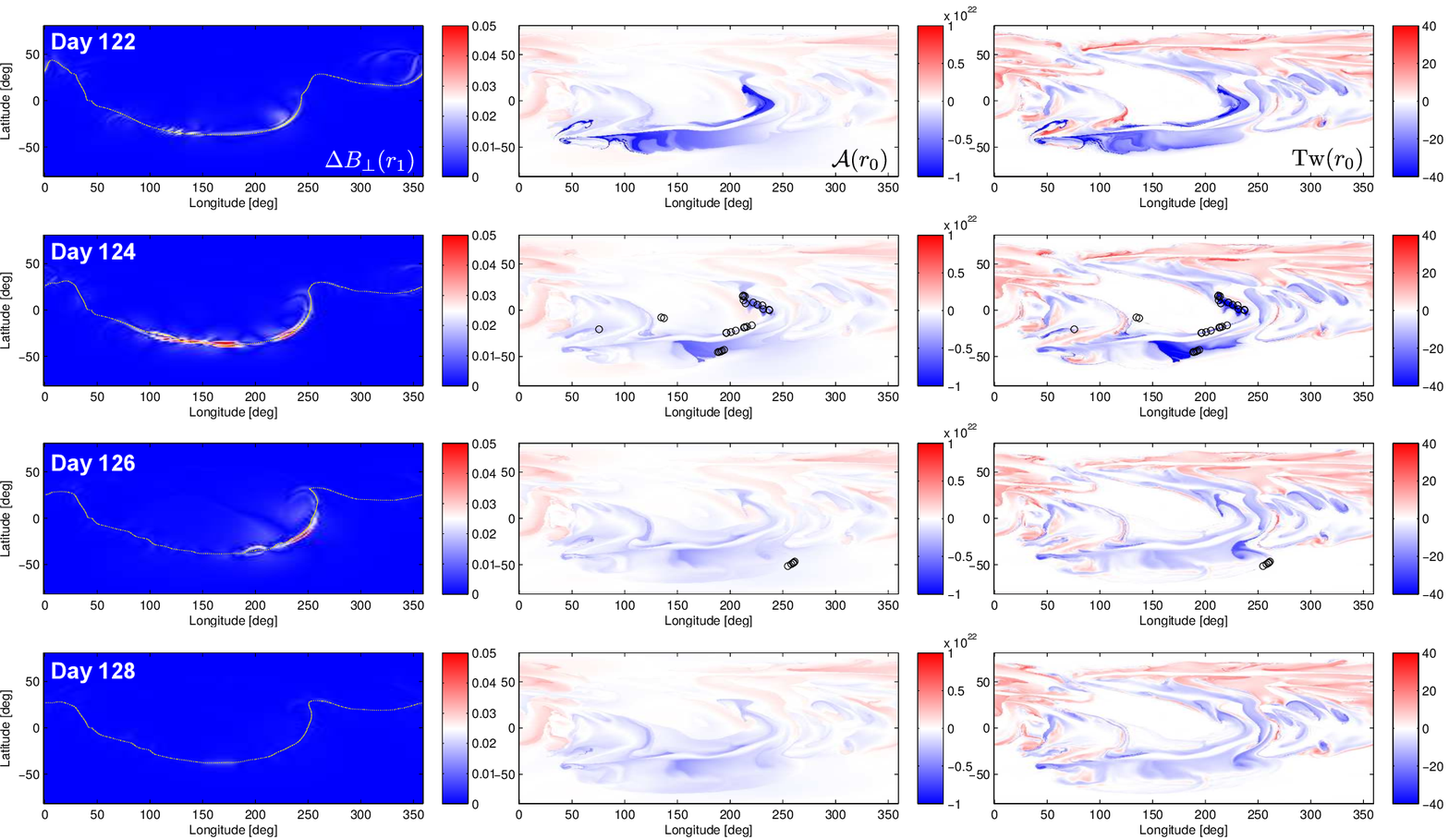}
\caption{Example of a flux rope ejection from period A. From top to bottom, the rows show days 122, 124, 126, and 128. The left column shows the (absolute) running daily difference of horizontal field $B_\perp := (B_\theta^2 + B_\phi^2)^{1/2}$ at the outer boundary $r=r_1$. The middle column shows the distribution of ${\cal A}$ on $r=r_0$ (saturated at $\pm 10^{22}\,\mathrm{Mx}$), and the right column shows the distribution of ${\rm Tw}$ at $r=r_0$ (saturated at $\pm 40$). The dashed lines in the left column show the neutral line where $B_r(r_1,\theta,\phi)=0$. Black circles in the other columns identify footpoints of field lines traced down from locations at $r=r_1$ where the running difference of $B_\perp$ exceeds $0.05\,\mathrm{G}\,\mathrm{day}^{-1}$. \textbf{Animated versions of this figure for both periods A and B are available in the online journal.}}
\label{fig:nonaxiropeA}
\end{figure*}

A significant feature of the evolution are the multiple flux rope ejections that occur over the 180-day simulations. These are visible as transient peaks in the open flux (Fig. \ref{fig:nonaxidiags}b), similar to the quadrupolar example (Fig. \ref{fig:quadrupolediags}) although less pronounced owing to their more localised nature. There are many more ejections in period B than in period A, due to the more complex corona in period B. The vertical grey lines in Fig. \ref{fig:nonaxidiags} indicate the times of significant flux rope ejections. These have been determined not from the open flux, but from monitoring the horizontal magnetic field at the outer boundary $r=r_1$. This is enhanced significantly during the ejection of flux ropes, as is shown by the left column of Fig. \ref{fig:nonaxiropeA}. This shows a running difference of $B_\perp := (B_\theta^2 + B_\phi^2)^{1/2}$ at $r=r_1$, for a particular ejection during period A. The ejection times shown in Fig. \ref{fig:nonaxidiags} were found by computing the number of grid points on each day with $B_\perp > 0.05\,\mathrm{G}\,\mathrm{day}^{-1}$, then identifying local maxima in this time series.

The second column of Fig. \ref{fig:nonaxiropeA} shows the distribution of ${\cal A}$ on the solar surface $r=r_0$. The black circles are the footpoints of field lines with $B_\perp > 0.05\,\mathrm{G}\,\mathrm{day}^{-1}$, so represent the footpoints of the ejected flux rope. It is clear that the region of strongest ${\cal A}$ is the erupting flux rope (and its overlying arcade). In fact, this flux rope is clearly seen in Fig. \ref{fig:nonaxifl2d}. Further evidence that these are the footpoints of the erupting rope comes from the significant weakening of ${\cal A}$ in this region following ejection of the rope. Similar behaviour is found for all of the ejections in periods A and B.

Finally, consider the right-most column of Fig. \ref{fig:nonaxiropeA}. This shows the dimensionless quantity
\begin{equation}
\mathrm{Tw} = \int_L\frac{\jb\cdot\Bb}{|\Bb|^2}\,\mathrm{d}l,
\end{equation}
which \citet{liu2016} call the twist number. In a force-free field, which is approximately the case in our model, we have $\jb = \alpha\Bb$ with $\alpha$ constant along each field line, so $\mathrm{Tw}$ is simply $\alpha\times\mathrm{length}(L)$. This measure is also an indicator of where twisted structures are located within the magnetic field, clearly identifying the erupting flux rope in Fig. \ref{fig:nonaxiropeA}. For this rope, ${\cal A}$ and $\mathrm{Tw}$ agree that this is the most significant twisted structure present, and agree on the sign of twist. But in general, $\mathrm{Tw}$ and ${\cal A}$ have different relative magnitude and sign. Partly, the difference in relative magnitude can be explained by the fact that ${\cal A}$ has units of magnetic flux while $\mathrm{Tw}$ is dimensionless. Put simply, a flux rope with the same field line curves but lower field strength would have the same $\mathrm{Tw}$, but weaker ${\cal A}$. This accounts for the lower values of ${\cal A}$ at high latitudes in Fig. \ref{fig:nonaxiropeA}, because this is a weak field region. However, significant differences in ${\cal A}$ and $\mathrm{Tw}$ also arise because $\mathrm{Tw}$ depends only on the local twist around a single field line, whereas ${\cal A}$ is a more global quantity. This tends to give ${\cal A}$ a smoother distribution within each magnetic subdomain, as is evident in Fig. \ref{fig:nonaxiropeA}. Overall, there is a significant correlation between ${\cal A}$ and $\mathrm{Tw}$, although the (rank) correlation coefficient is only about $0.6$. (This value remains steady after an initial transient phase of about 24 days where the correlation is lower.) Perhaps the most compelling reason to use ${\cal A}$ rather than $\mathrm{Tw}$ is that $\mathrm{Tw}$ is not an ideal invariant \citep[see][]{moffatt1992,berger2006}.

\section{Conclusion} \label{sec:conc}

We have shown how field line helicity ${\cal A}$ is an invaluable tool for quantifying the distribution of topological structure within the Sun's corona. It is straightforward to compute from a 3D magnetic field, by first computing an appropriate vector potential. It is a physically meaningful measure representing the linkage of magnetic flux around each magnetic field line in the domain. In particular, it is invariant under ideal motions \edit{within the domain, provided that the field line footpoints on the boundary remain fixed.}

\edit{Although the value of ${\cal A}$ for a given field line is computed by integrating $\Ab$ along that single field line, the vector potential $\Ab$ being integrated is fundamentally a non-local quantity. This enables ${\cal A}$ to measure the linking with other magnetic field lines, but it does mean that knowledge of the wider magnetic field is required even to compute ${\cal A}$ on a single field line.}

\edit{We mentioned, in Section \ref{sec:flh}, that ${\cal A}$ is a meaningful density for the total magnetic helicity, being the limiting helicity on a infinitesimal tubular domain around each magnetic field line. In fact, this is the finest possible decomposition of magnetic helicity into subdomains that will remain ideal invariants. Any finer decomposition would necessarily have interfacial surfaces in the corona with $\Bb\cdot\nb\neq 0$, across which there would be helicity fluxes even in an ideal evolution.}

Although the decomposition into field lines has an infinite number of subdomains, we have seen (e.g., Fig. \ref{fig:nonaxiropeA}) that the distribution of ${\cal A}$ tends to be rather smooth, on account of its non-local definition. One could therefore give a first-order characterisation of the magnetic structure by integrating ${\cal A}$ over discrete topological subdomains, following decomposition of the magnetic skeleton \citep[e.g.,][]{haynes2010}. We have not pursued this idea here, as identifying the skeleton is computationally challenging in non-potential fields \citep[cf.][]{edwards2015}. However, a similar idea was proposed by \citet{longcope2008}, who defined the ``additive self-helicity'' of a sub-domain. Computation for simulations of a twisted magnetic flux tube were able to relate this quantity to the stability of the flux tube \citep{malanushenko2009}.

The gauge dependence of ${\cal A}$ arises purely from the fact that coronal magnetic field lines end on the boundaries rather than being closed loops. This gauge dependence is unavoidable; however, we have shown that every gauge is physically meaningful, corresponding to a different definition of what it means for flux to be linked with a magnetic loop. We have suggested that the DeVore gauge is a practical choice where not only is $\Ab$ is easy to compute, but the resulting field line helicity is appropriate for measuring twisted structures forming in the lower corona. An alternative way to choose a gauge would be to fix a vector potential where $\Ab\times{\boldsymbol n}$ matches some chosen reference field on the boundary, as in the commonly-used relative helicity \citep{berger1984a}. But really the choice of reference field is just another way of viewing the choice of gauge \citep{prior2014}. 

Having shown that field line helicity is a useful tool for coronal simulations, there are many possible future applications. An obvious one is to try to identify the locations where flux rope eruptions will occur, but there are many others. For example, we have, in this paper, neglected the direct emergence of already-twisted structures from the solar interior, and we have also neglected the net injection of helicity by small-scale convective motions. The relative importance of these two effects compared to surface shearing is important to establish, particularly for explaining the hemispheric pattern of helical structures in the corona \citep{pevtsov2003}. It will also be needed in order to make improved estimates of the Sun's helicity output over the solar cycle \citep[cf.][]{devore2000}. Another application will be to compare different methods of simulating the coronal magnetic field evolution -- for example, how accurate is the magneto-frictional approximation? What is the importance of including thermodynamics? What is the effect of different parametrizations of turbulent diffusion in the corona? Or how best do we drive coronal simulations based on limited photospheric data \citep[e.g.,][]{kazachenko2014}? We hope to address some of these questions in future research.

\begin{acknowledgements}
This work was supported by STFC consortium grant ST/K001043/1 to the universities of Dundee and Durham. ARY also thanks the US Air Force Office of Scientific Research for support through a grant from the Basic Research Initiative ``Understanding the interaction of CMEs with the solar-terrestrial environment''. We are grateful to Carl Henney for supplying the ADAPT maps for Sect. \ref{sec:nonaxi}, and thank Alexander Russell, Christopher Prior, and Mitchell Berger for useful discussions.
\end{acknowledgements}

\bibliographystyle{aa}
\bibliography{yeates}

\appendix
\section{Existence of a surface with flux ${\cal A}$}

Consider the magnetic field line $L_2$ in Fig. \ref{fig:flux}, whose endpoints $x_1$ and $x_2$ both lie on the boundary $r=r_0$. We will show that there exists a curve $\gamma$ from $x_1$ to $x_2$, lying on the surface $r=r_0$, such that
\begin{equation}
\int_\gamma\Ab\cdot\,\mathrm{d}\lb = 0.
\end{equation}
This means that the surface bounded by $L$ and $\gamma$ has flux equal to ${\cal A}(L)$.

\begin{figure}
\resizebox{\hsize}{!}{\includegraphics{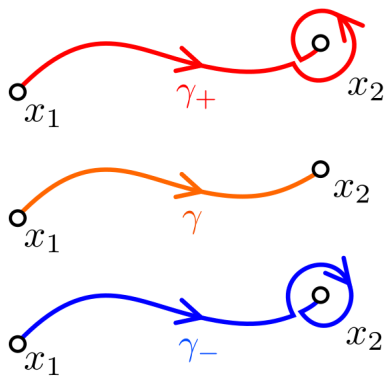}}
\caption{The original and deformed curves, all on the boundary $r=r_0$.}
\label{fig:gamma}
\end{figure}

To see that such a curve exists, suppose that we continuously deform the original curve $\gamma$ into either $\gamma_+$ or $\gamma_-$, as shown in Fig. \ref{fig:gamma}. Since $x_2$ is a field line footpoint, we must have $B_r(x_2)\neq 0$. If  $B_r(x_2)>0$, then the curve $\gamma_+$ will have a larger value of $\int \Ab\cdot\,\mathrm{d}\lb$ than $\gamma$, and the curve $\gamma_-$ will have a smaller value. By further deforming these curves to encircle $x_2$ more than once, we may ensure that $\int_{\gamma_+}\Ab\cdot\,\mathrm{d}\lb > 0$ and $\int_{\gamma_-}\Ab\cdot\,\mathrm{d}\lb < 0$. By continuity, there must exist some intermediate curve with vanishing integral.

It is easy to see that there are many such curves $\gamma$ with the required property, for any pair of footpoints $x_1$, $x_2$, and even if the gauge of $\Ab$ is fixed. But all of the corresponding surfaces will have the same flux ${\cal A}(L)$, and this will be an ideal invariant if footpoint motions are disallowed. 

Clearly this argument applies equally if both footpoints lie on $r=r_1$ (a rarer situation in the corona). But what about an open field line, where $x_1$ lies on $r=r_0$ and $x_2$ on $r=r_1$? Now the curve $\gamma$ that completes the loop must pass through $D$, rather than lying on the boundary. But, provided this portion of $\gamma$ is chosen to be a magnetic field line, the resulting surface will again have an ideal-invariant flux. Again, this can be made equal to ${\cal A}(L)$ (the field line helicity of the original field line) by appropriately choosing the portions of $\gamma$ on the two boundaries. So ${\cal A}$ still represents an ideal-invariant flux, even if the field line is open.

\end{document}